\documentclass[12pt]{iopart}

\usepackage{graphicx}
\usepackage [latin1]{inputenc}
\usepackage{cite}
\usepackage{xcolor}
\usepackage[english]{babel}
\usepackage{hyperref}
\usepackage{dirtytalk}
\usepackage{amsmath}
\usepackage{amssymb}
\usepackage[letterpaper,top=2cm,bottom=2cm,left=2cm,right=2cm,marginparwidth=1.75cm]{geometry}

\usepackage{stmaryrd}

\newcommand{\be}{\begin{equation}}
\newcommand{\ee}{\end{equation}}

\begin{document}


\title[Reducing turbulent transport combining intrinsic rotation and LMD regime]{Reducing turbulent transport in tokamaks by combining intrinsic rotation and the low momentum diffusivity regime}

\author[]{Haomin Sun$^{1,*}$, Justin Ball$^{1}$, Stephan Brunner$^{1}$, Anthony Field$^{2}$, Bhavin Patel$^{2}$, Daniel Kennedy$^{2}$, Colin Roach$^{2}$, Diego Jose Cruz-Zabala$^{3}$, Fernando Puentes Del Pozo$^{3}$, Eleonora Viezzer$^{3}$, and Manuel Garcia Munoz$^{3}$}

\address{$^1$Ecole Polytechnique F\'ed\'erale de Lausanne (EPFL), Swiss Plasma Center (SPC), CH-1015 Lausanne, Switzerland}
\address{$^2$UKAEA (United Kingdom Atomic Energy Authority), Culham Campus, Abingdon, Oxfordshire, OX14 3DB, UK}
\address{$^3$Department of Atomic, Molecular and Nuclear Physics, University of Seville, Seville, Spain}
\ead{$^{*}$haomin.sun@epfl.ch}
\vspace{10pt}
\begin{indented}
\item[] December 2024
\end{indented}


\vspace{10pt}


\date{\today}

\begin{abstract}
Based on the analysis of a large number of high-fidelity nonlinear gyrokinetic simulations, we propose a novel strategy to improve confinement in \textcolor{black}{spherical} tokamak plasmas by combining up-down asymmetric flux surface shaping with the Low Momentum Diffusivity (LMD) regime. We show that the intrinsic momentum flux driven by up-down asymmetry creates strong flow shear in the LMD regime that can significantly reduce energy transport, increasing the critical gradient by up to $25\%$. In contrast to traditional methods for generating flow shear, such as neutral beam injection, this approach requires no external momentum source and is expected to scale well to large fusion devices. The experimental applicability of this strategy in spherical tokamaks is addressed via simulations by considering actual equilibria from MAST and a preliminary equilibrium from SMART.


\end{abstract}

\maketitle


\textbf{\textit{Introduction--}}It is well known that strong toroidal rotation (specifically $E\times B$ rotation shear) can reduce turbulent transport \cite{Biglari1990,Stambaugh1990enhanced,JETL-GEriksson_1997,Angioni2001IntrinsicRotation,Peeters2005LinearToroidal,deVries_2008,schekochihin2008,JETMantica2009PRL,Ida2009PRL,NewtonFlowShearUnderstanding2010,Highcock2011POP,BarnesFlowShear2011,highcock2012,schekochihin2012,ChristenFlowShear2018,ben2019,ball2019,JETJ.M.Noterdaeme_2003,JETdeVries_2006} and stabilize MagnetoHydroDynamic (MHD) instabilities \cite{RotationMHDGarofalo2002PRL,RotationMHDAiba2009NF,RotationMHDAiba2011NF,RotationMHDChu1999NF,RotationMHDWahlberg2000POP}, thereby improving the prospects for tokamak fusion power plants. Traditional methods of driving plasma rotation such as Neutral Beam Injection (NBI) \cite{Groebner1990NBIrotation,Suckewer1981NBIrotation,Goumiri2016NSTXNBI} or Radio Frequency (RF) waves \cite{Hsuan1996ICHrotation,Chang1999ICRH,Chan2002RFrotation,Li_2011EASTRF,Lyu2020RFrotation} are not expected to scale well to large devices \cite{YueqiangLiu_2004}. Taking NBI as an example, we see that the ratio of the momentum $p_{\text{NBI}}$ over the energy $E_{\text{NBI}}$ injected by the beam scales as $p_{\text{NBI}}/E_{\text{NBI}}\sim 1/\sqrt{E_{\text{NBI}}}$ \cite{Parra_2011_PRL_Momentum_optimum}. Thus, the more energetic beams needed to penetrate into larger and denser plasmas will be less effective at driving rotation, making externally driven rotation increasingly difficult for large devices.

An intriguing alternative is to utilize intrinsic rotation \cite{Camenen2010MomentumTransport,Hornsby2017momentum,zhu2024intrinsic}, spontaneously driven by the plasma turbulence itself under certain conditions. The symmetry properties of gyrokinetics \cite{Peeters2005LinearToroidal,ParraUpDownSym2011,Peeters_2011NFReview,Parratheory_2015} show that the only way to create flow of the order of the sound speed in the tokamak core is to break the up-down symmetry of magnetic flux surfaces about the midplane \cite{Camenen2009TUBTrans,ball2014}. This approach is independent of external sources, \textcolor{black}{has been demonstrated experimentally \cite{Camenen2010MomentumTransport}} and is expected to scale well to large devices (as it persists in the limit $\rho^*=\rho_i/a\to 0$, where $\rho_i$ is ion gyroradius and $a$ is tokamak minor radius). One can calculate the radial profile of rotation in the steady-state operation of a tokamak by requiring the total toroidal angular momentum flux to be zero, which is well approximated by considering only the ion contribution ($\Pi_i=0$) due to their large mass compared to electrons. 
Since the momentum flux depends on both the toroidal ion rotation $\Omega_i$ and its radial derivative $d\Omega_i/dx$, it is common to Taylor expand $\Pi_i$ about $\Omega_i=0$ and $d\Omega_i/dx=0$ to find \cite{ball2014}
\begin{equation}
\Pi_i\left(\Omega_i,\frac{d\Omega_i}{dx}\right)\simeq\Pi_{i,int}
-n_im_iR^2_0D_{\Pi i}\frac{d\Omega_i}{dx}-n_im_iR^2_0P_{\Pi i}\Omega_i=0,
\label{eq_momentum}
\end{equation}
where $\Pi_{i,int}=\Pi_i(0,0)$ is the intrinsic momentum flux generated by up-down asymmetry and the second and third terms are the diffusive and pinch terms, respectively. Here, $n_i$ is ion density, $m_i$ is the ion mass, $R_0$ is the major radius of the tokamak, $x$ is the radial coordinate, $D_{\Pi i}$ is the angular momentum diffusion coefficient and $P_{\Pi i}$ is the angular momentum pinch coefficient. Since the pinch term is typically smaller than the diffusive term \cite{Peeters2007PRLpinchterm,Guttenfelder_2017,Zimmermann_2022,Zimmermann_2023}, we neglect it for now, but show in Ref. \cite{sun2024physicslowmomentumdiffusivity} that this will only cause an \textit{underestimation} of the flow shear created by up-down asymmetry. The strength of the ion momentum diffusivity is typically compared with the ion heat diffusivity $D_{Q i}$ through the ion Prandtl number
\begin{equation}\label{Prandtl1}
    \text{Pr}_i=\frac{D_{\Pi i}}{D_{Q i}}.
\end{equation}
This allows one to estimate the rotation shear being driven by a given turbulence level. A low Prandtl number means that a given momentum source (external or intrinsic) drives larger rotation shear at a given turbulence level. Without considering the pinch term, Eq. \eqref{eq_momentum} indicates that maximizing $\Pi_{i,int}$ or minimizing ion momentum diffusivity will strengthen velocity shear. Maximizing $\Pi_{i,int}$ has been the subject of previous work \cite{ball2018}. The focus of this letter is to minimize the Prandtl number so as to help stabilize turbulence. In conventional aspect ratio tokamaks, $\text{Pr}_i\simeq 1$ has been experimentally observed \cite{Tala_2011,Weisen_2012}, motivating many theoretical works to assume this \cite{Diamond2008momentumtransport,Diamond_2009,ball2014,ball2018}.
Recently, however, it has been shown by gyrokinetic simulations that a significantly lower Prandtl number can be achieved \cite{Holod2008GKSim,Camenen_2011NFReview}, especially at tight aspect ratio and low safety factor \cite{casson2009,Ben_2015_intrinsic,Ben_2019}. However, the Prandtl number calculations in many works were simplified, since either \textit{parallel} momentum flux \cite{casson2009} or the toroidal angular momentum flux generated by flow shear \textit{parallel} to the magnetic field \cite{Ben_2015_intrinsic,Ben_2019} was calculated. A more complete calculation must consider all components of \textit{toroidal} angular momentum flux \cite{ParraUpDownSym2011,BarnesFlowShear2011} generated by \textit{toroidal} flow shear as actually present in tokamaks \cite{AbelGyrokineticsDeriv2012}. A thorough scan of the dependence of the Prandtl number on the various geometric properties of tight aspect ratio tokamaks has not been performed, nor has a self-consistent study of the stabilization effect of intrinsic flow shear by up-down asymmetry on plasma turbulence. Despite recent experimental tokamak data showing that one can obtain a Prandtl number lower than one \cite{deVries_2008,McDermott_2011,BuchholzPrandtl2015,Guttenfelder_2017,Hornsby_2018,Zimmermann_2022,Zimmermann_2023}, these results are often hard to interpret given the challenge of separating the three terms in Eq. \eqref{eq_momentum}, especially for tight aspect ratio spherical tokamaks \cite{Meyer_2009,Zimmermann_2022}. All these theoretical and experimental issues call for a comprehensive theoretical study of toroidal angular momentum transport and how this can be combined with rotation drive mechanisms to stabilize turbulence. 

We perform a thorough study of the Low Momentum Diffusivity (LMD) regime using a large number of high-fidelity nonlinear (NL) local gyrokinetic simulations. We first use a circular geometry to determine the parameter regime that achieves LMD. We then consider a tilted elliptical geometry in order to drive rotation from up-down asymmetry. By scanning the flow shear $d\Omega_i/dx$ to find the value at which momentum flux is zero, we show that turbulence can self-consistently drive an intrinsic flow shear that significantly reduces the heat flux. The experimental applicability of this approach is addressed via simulations of a MAST tokamak equilibrium and a preliminary SMART equilibrium. To our knowledge, this work not only provides the first comprehensive numerical study on toroidal angular momentum diffusivity in the LMD regime, but also demonstrates the first potentially practical method of driving strong rotation shear in future tokamak power plants.

\textbf{\textit{Methods--}}We use the well-benchmarked code GENE \cite{JenkoGENE2000,GoerlerGENE2011} to perform a large number of flux tube (local) gyrokinetic simulations and model core plasma turbulence. We consider the Miller representation \cite{Millergeometry1998} to parameterize the flux surfaces. To make our analysis as accurate as possible, we consider the actual \textit{toroidal} flow shear (including its parallel and $E\times B$ components) to drive momentum flux as well as calculate the exact \textit{toroidal} angular momentum flux (according to Eq. (A.2) in Ref. \cite{Sun2024NF}). 

\begin{figure}
    \centering
    \includegraphics[width=0.49\textwidth]{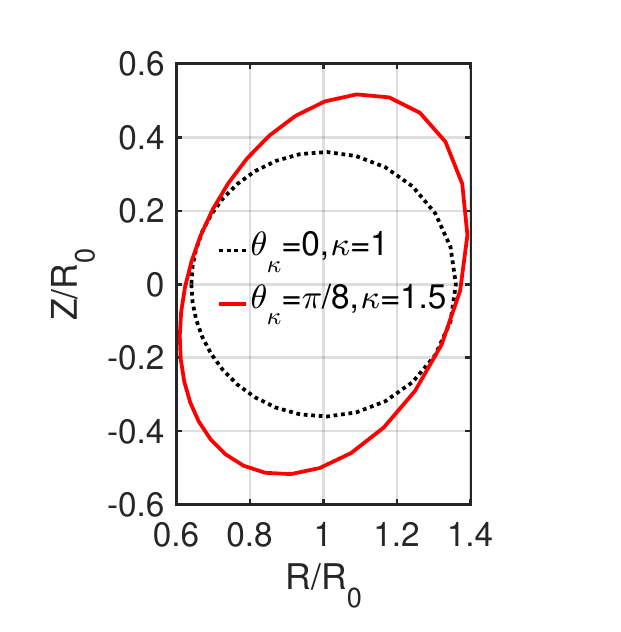}
    \caption{\textcolor{black}{The figure shows the poloidal cross-sectional shape of the flux surfaces of circular (black dotted), and tilted elliptical (red solid) flux surfaces that we use in gyrokinetic simulations, where $\kappa$ is elongation, $\theta_{\kappa}$ is the elongation tilt angle, $R$ and $Z$ are horizontal and vertical axes, respectively.}}
    \label{fig3updownasyfluxsurface}
\end{figure}
\textbf{\textit{Analysis--}}We start with circular concentric flux surfaces (black dotted line in Fig. \ref{fig3updownasyfluxsurface}), an inverse aspect ratio $\epsilon=r_0/R_0=0.36$ ($r_0$ is the minor radial location of the flux tube), adiabatic electrons, equal electron and ion temperatures, and a density gradient of $R_0/L_n=2.22$. These parameters are the same as the Cyclone Base Case (CBC) \cite{DimitsShift2000} (except the tighter aspect ratio) and produces Ion Temperature Gradient (ITG) driven turbulence. 
We perform a parameter scan over safety factor $1.05\leq q\leq 4.55$, magnetic shear $0.1\leq\hat{s}\leq 1.6$ and ion temperature gradient $4.96\leq R_0/L_{Ti}\leq 12.96$, comprising a 3D data set with 576 NL gyrokinetic simulations, where $L_n$ and $L_T$ are the characteristic gradient scale lengths of density and temperature, respectively. To drive non-zero toroidal angular momentum flux and compute the momentum diffusivity, we impose toroidal flow shear $d\Omega_i/dx$ including a parallel and a perpendicular component of $\omega_{\perp}=-(r_0/q)d\Omega_i/dx=0.12c_s/R_0$, where $c_s=\sqrt{T_e/m_i}$ is the sound speed, and $T_e$ is electron temperature. We use a simulation box with widths $L_x\times L_y=200\rho_i\times 200\rho_i$ and number of grid points $(n_{k_x},n_{k_y},n_z,n_{v_{||}},n_{\mu})=(192,64,64,32,9)$, where $\rho_i$ is the ion gyroradius, $(x,y,z,v_{||},\mu)$ are the \text{radial}, \text{binormal}, \text{parallel}, \text{parallel velocity}, and magnetic moment directions, respectively, $k_x$ and $k_y$ are the wavevectors in $x$ and $y$ directions. The heat flux and toroidal angular momentum flux, $\hat{Q}_i$ and $\hat{\Pi}_i$, are respectively normalized to gyroBohm units $c_s n_i T_i(\rho_i/R_0)^2$ and $c^2_s m_i n_i R_0 (\rho_i/R_0)^2$, where $T_i$ is the ion temperature.

\begin{figure}
    \centering
    \includegraphics[width=0.94\textwidth]{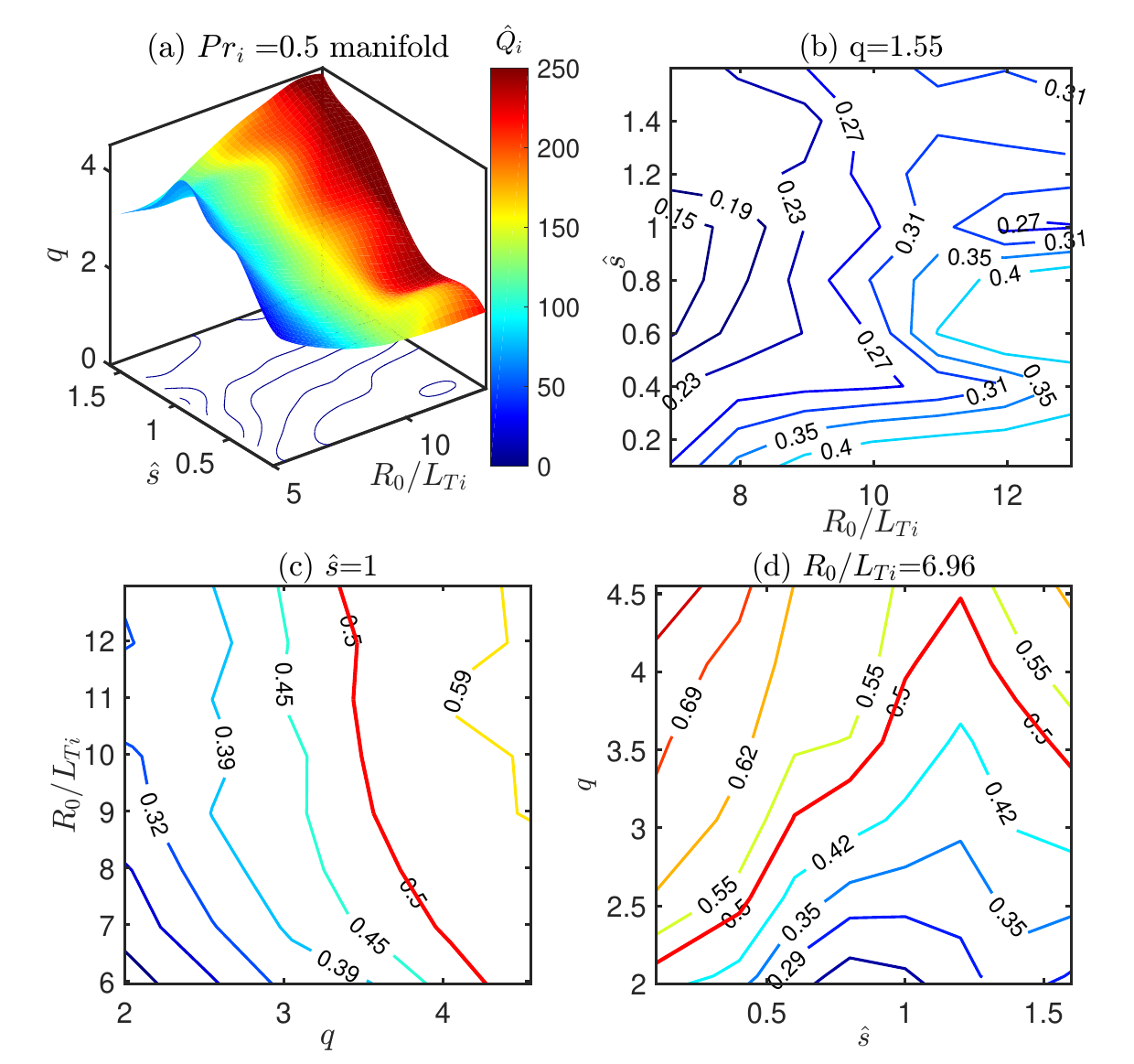}
    \caption{Visualizations of the Prandtl number. (a) shows the surface with $\text{Pr}_i=0.5$, below which we define as the LMD regime. The color map on the surface denotes the heat flux $\hat{Q}_i$. The contours in the bottom plane represent lines of constant $q$ values on the surface. (b)-(d) provide contours of Prandtl number fixing one of our three parameters: (b) $q=1.55$, (c) $\hat{s}=1.0$, and (d) $R_0/L_{Ti}=6.96$. The red contour lines denote $\text{Pr}_i=0.5$.}
    \label{fig1_manifoldcontour}
\end{figure}

Figure \ref{fig1_manifoldcontour} (a) shows the constant Prandtl number manifold $\text{Pr}_i=0.5$ that we construct by linearly interpolating the $\text{Pr}_i$ data from our 3D parameter scan. In simulations with up-down symmetric geometries, the Prandtl number is calculated by 
\begin{equation}\label{eq_Prandtlnumber}
\text{Pr}_i=\frac{\hat{\Pi}_i}{\hat{Q}_i}\frac{R_0}{L_{Ti}}\frac{\epsilon}{q\,\omega_{\perp}}\frac{c_s}{R_0}.
\end{equation}
Below the manifold, corresponding to lower values of $q$ at fixed $\hat{s}$ and $R_0/L_{Ti}$, $\text{Pr}_i$ is smaller than $0.5$. \textcolor{black}{We choose to define the region with $\text{Pr}_{i} \lesssim 0.5$ as the LMD regime. Having an objective criterion for ``low momentum diffusivity'' will be useful in discussing the properties of the regime, but we emphasize that the momentum diffusivity changes gradually and nothing distinct occurs at $\text{Pr}_{i} = 0.5$. The reason for choosing this number is because this is about the lowest value of Prandtl number measured in previous experiments of large aspect ratio tokamaks \cite{Zimmermann_2023}. } Only the nonlinearly unstable region of the parameter space is presented in the figure. We see that $\hat{s}\simeq 1$ and low $R_0/L_{Ti}$ favor a low $\text{Pr}_i$, reflected by a larger range in $q$ with low $\text{Pr}_i$. The heat flux shown by the color of the manifold changes primarily with $R_0/L_{Ti}$, \textcolor{black}{indicating that $q$ and $\hat{s}$ do not affect $\hat{Q}_i$ as significantly as $R_0/L_{Ti}$}. To show the parameter dependence of the Prandtl number more clearly, several contours of $\text{Pr}_i$ are shown in Fig. \ref{fig1_manifoldcontour} (b)-(d) by holding either $q$, $\hat{s}$ or $R_0/L_{Ti}$ constant. It is clear that $\text{Pr}_i$ has a non-trivial dependence on these parameters. Figure \ref{fig1_manifoldcontour} (c) and (d) show that $\text{Pr}_i$ depends strongly on $q$, and Fig. \ref{fig1_manifoldcontour} (b) shows that at low $q$, $\text{Pr}_i$ becomes small over a large range in $\hat{s}$ and $R_0/L_{Ti}$. This feature, together with a small aspect ratio $1/\epsilon=2.78$ suggests that a more poloidally angled magnetic field reduces $\text{Pr}_i$. While the dependence of $\text{Pr}_i$ on $R_0/L_{Ti}$ is weaker, being closer to marginal stability reduces $\text{Pr}_i$. \textcolor{black}{The Prandtl number and LMD regime depend strongly on magnetic shear, as it is a factor of $\sim 2$ higher at a low value of $\hat{s}=0.1$ compared to $\hat{s}\simeq 1$, shown in Fig. \ref{fig1_manifoldcontour} (d).} As shown in our companion paper \cite{sun2024physicslowmomentumdiffusivity}, only considering the \textit{parallel} component of $\hat{\Pi}_i$ results in an overestimate of $\text{Pr}_i$. Adding kinetic electrons retains the above features in $\text{Pr}_i$, while increasing the aspect ratio $1/\epsilon$ strongly increases $\text{Pr}_i$ \cite{sun2024physicslowmomentumdiffusivity}. \textcolor{black}{Therefore, spherical tokamaks are best able to experimentally validate the findings in this paper. Due to the small mass of electrons, electron-scale turbulence is expected to contribute little to momentum transport. Therefore, if one could stabilize the ion-scale turbulence, the momentum diffusivity should become significantly smaller, making it easier to achieve fast rotation. In our companion paper \cite{sun2024physicslowmomentumdiffusivity}, we also show that changing to Trapped Electron Mode (TEM) turbulence does not affect the Prandtl number significantly. The effect of electromagnetic turbulence on the Prandtl number, e.g. Micro-Tearing Modes (MTMs) or Kinetic Ballooning Modes (KBMs), is left for future investigations. }



\begin{figure}
    \centering
    \includegraphics[width=0.94\textwidth]{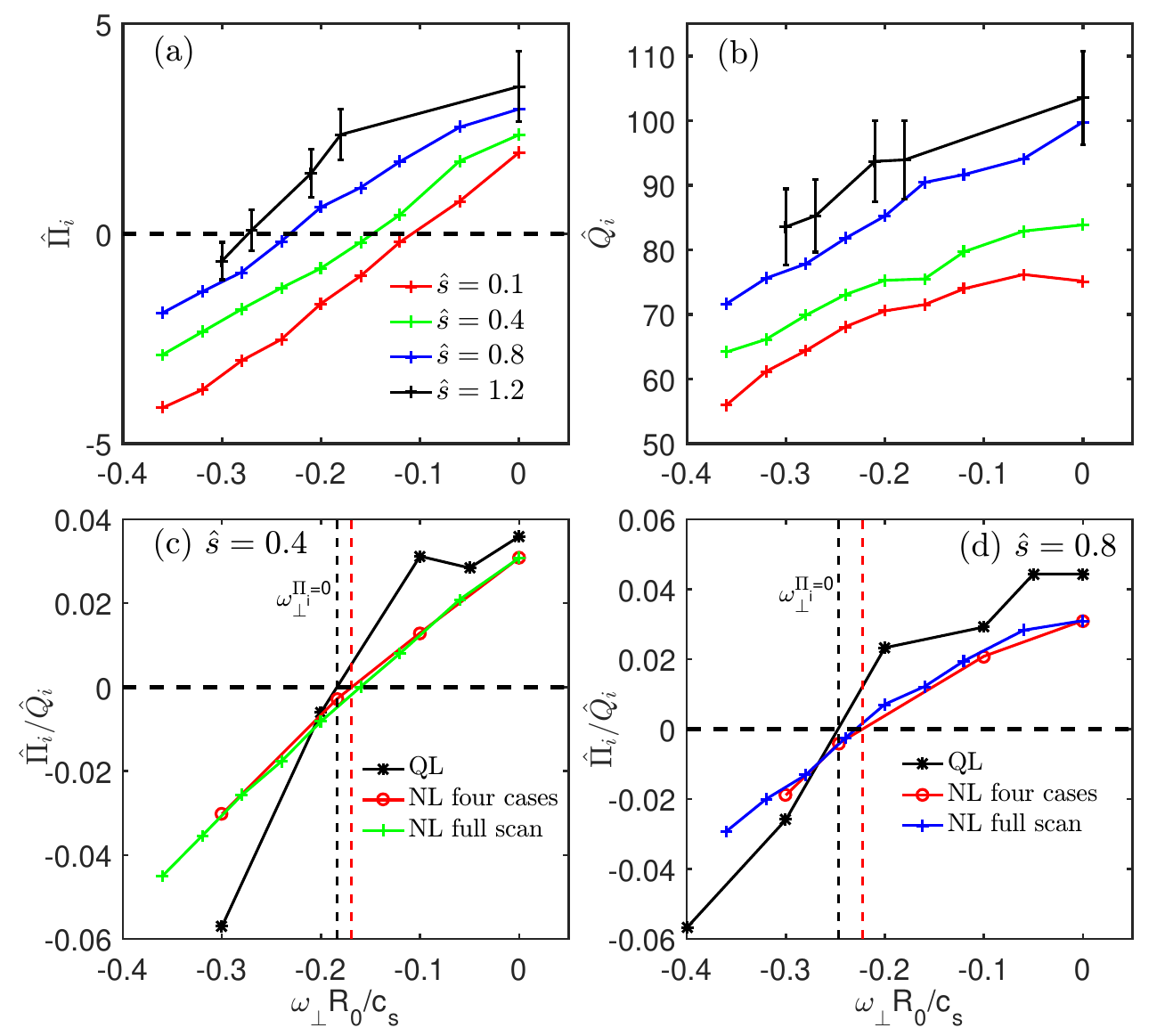}
    \caption{Subplots (a) and (b) show $\hat{\Pi}_i$ and $\hat{Q}_i$ as a function of $\omega_{\perp}$ obtained with NL simulations for $\hat{s}=0.1$ (red), $\hat{s}=0.4$ (green), $\hat{s}=0.8$ (blue) and $\hat{s}=1.2$ (black). Subplots (c) and (d) show a comparison of $\hat{\Pi}_i/\hat{Q}_i$ as a function of $\omega_{\perp}$ between the QL model (black), NL simulations with four cases of $\omega_{\perp}=\bigl\{0,-0.1,-0.3,\omega^{{\Pi}_i=0}_{\perp}\bigl\}c_s/R_0$ (red), and the full NL scans shown in (a) and (b) for $\hat{s}=0.4$ (green) and $\hat{s}=0.8$ (blue), respectively. All results are for $q=2.05$, $R_0/L_{Ti}=10.96$. The error bars (calculated by a rolling average on the time traces) are shown for some example cases.}
    \label{fig2_flowshearscan}
\end{figure}
Next, we combine LMD with a source of intrinsic momentum flux to drive strong flow shear. To predict the self-consistent rotation gradient that could be achieved in experiments, we perform NL gyrokinetic simulations with adiabatic electrons for tilted elliptical magnetic flux surfaces (see red solid line in Fig. \ref{fig3updownasyfluxsurface}) with $\epsilon=0.36$, elongation $\kappa=1.5$, and elongation tilt angle $\theta_{\kappa}=\pi/8$. This tilt angle was identified to generate the strongest intrinsic momentum flux in a CBC-like equilibrium \cite{ball2018}. 

Without external momentum sources, steady state operation of a tokamak requires $\hat{\Pi}_i=0$. To enforce this for a given equilibrium in local gyrokinetic simulations, we need to scan over the flow shear $\omega_{\perp}$ to find the value at which $\hat{\Pi}_i=0$. This is illustrated in Fig. \ref{fig2_flowshearscan} (a), where we see that as we increase the magnitude of the (negatively valued) flow shear, $\hat{\Pi}_i$ gradually drops to zero and then changes sign. The flow shear for which $\hat{\Pi}_i=0$ is the expected steady-state value in experiment. We see that a higher $\hat{s}$ results in a higher magnitude of the steady-state flow shear $|\omega_{\perp}|$. This is both because higher $\hat{s}$ lowers the Prandtl number (reflected in Fig. \ref{fig1_manifoldcontour} (d)) and because higher $\hat{s}$ increases the intrinsic momentum flux (reflected by values of $\hat{\Pi}_i$ at $\omega_{\perp}=0$ in Fig. \ref{fig2_flowshearscan} (a)) \cite{Parratheory_2015,ball2018}. As expected, the heat flux is reduced by a larger $|\omega_{\perp}|$ \cite{casson2009} as shown in Fig. \ref{fig2_flowshearscan} (b). \textcolor{black}{We also see that, in line with existing literature \cite{Kinsey2006popshateffect}, reducing $\hat{s}$ directly reduces the heat flux. Unfortunately, achieving low $\hat{s}$ over a broad radial region is challenging in experiments. }

Making a 4D NL scan of $(q,\hat{s},R_0/L_{Ti},\omega_{\perp})$ is computationally intensive. To reduce the numerical costs, we use a QuasiLinear (QL) model developed specifically to estimate $\hat{\Pi}_i/\hat{Q}_i$ \cite{Sun2024NF}. This allows us to scan $\omega_{\perp}$ and find the steady-state flow shear $\omega^{{\Pi}_i=0}_{\perp}$ such that $\hat{\Pi}_i(\omega^{\Pi_i=0}_{\perp})=0$. We then compare two NL simulations, one with $\omega_{\perp}=\omega^{\Pi_i=0}_{\perp}$ and one with $\omega_{\perp}=0$, to determine the heat flux reduction by intrinsic rotation. A benchmark of the flux ratio $\hat{\Pi}_i/\hat{Q}_i$ between NL simulations and our QL model is shown in Fig. \ref{fig2_flowshearscan} (c) and (d). We see that our QL model predicts $\omega^{{\Pi}_i=0}_{\perp}$ well. Note that we use a linear interpolation between neighboring data points. To estimate the Prandtl number for up-down asymmetric cases, we perform two NL simulations at $\omega_{\perp}=\bigl\{-0.1,-0.3\bigl\}c_s/R_0$. As explained in Ref. \cite{sun2024physicslowmomentumdiffusivity}, we adjust Eq. \eqref{eq_Prandtlnumber} to be
\begin{equation}\label{eq_Prandtlnumber22}
\overline{\text{Pr}}_i=\Delta\left(\frac{\hat{\Pi}_i}{\hat{Q}_i}\right)\frac{R_0}{L_{Ti}}\frac{\epsilon}{q\Delta\omega_{\perp}}\frac{c_s}{R_0},
\end{equation}
so that the Prandtl number is proportional to the slope in Fig. \ref{fig2_flowshearscan} (c) and (d), where $\Delta$ indicates the difference of quantities between their values at $\omega_{\perp}=-0.1c_s/R_0$ and $-0.3c_s/R_0$. Hence, four NL simulations are performed for each $(q,\hat{s},R_0/L_{Ti})$, i.e. $\omega_{\perp}=\bigl\{0,-0.1,-0.3,\omega^{{\Pi}_i=0}_{\perp}\bigl\}c_s/R_0$ (shown in Fig. \ref{fig2_flowshearscan} (c) and (d) by the red lines). Note that the QL model is only used to obtain $\omega^{{\Pi}_i=0}_{\perp}$.

\begin{figure}
    \centering
    \includegraphics[width=0.94\textwidth]{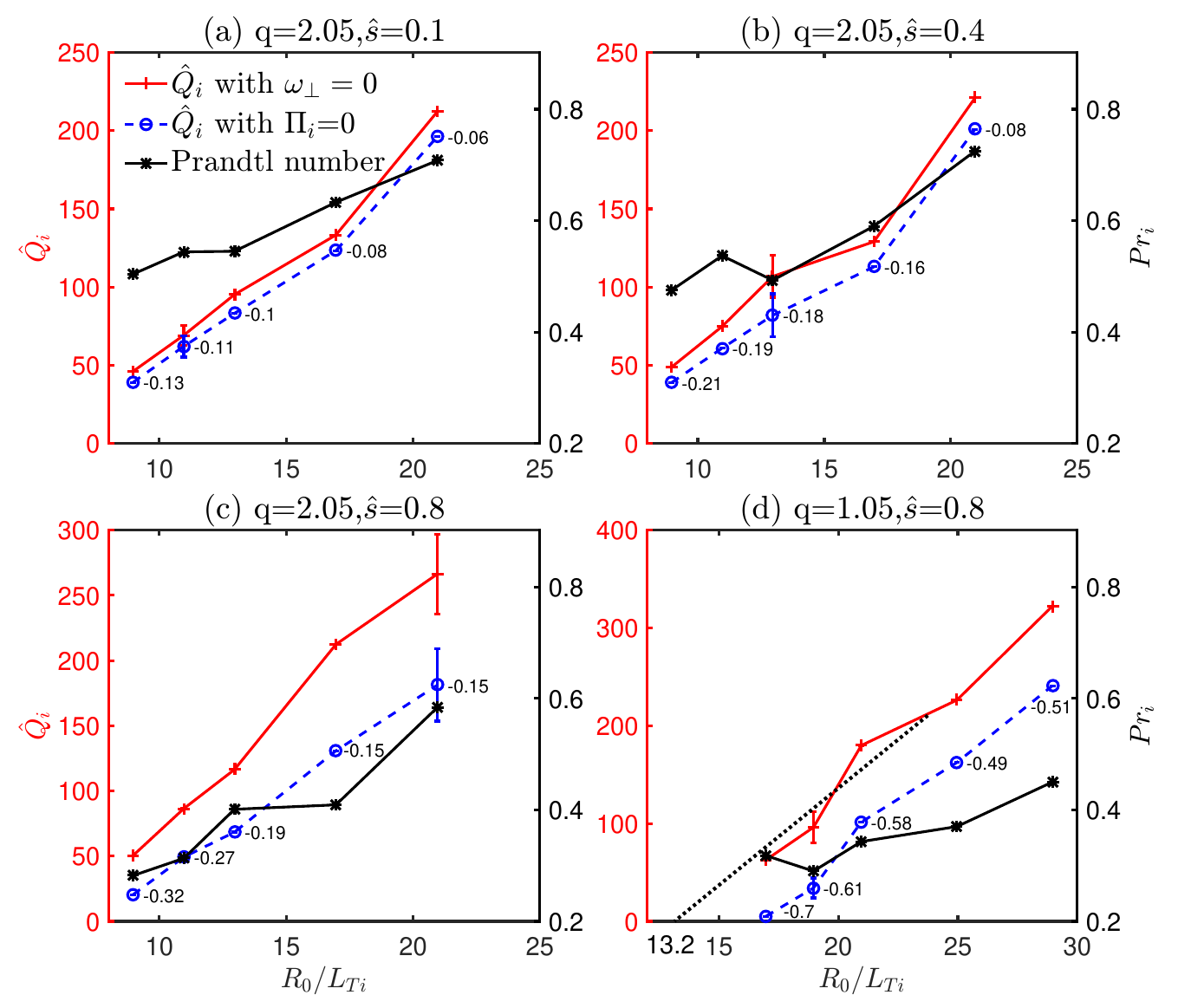}
    \caption{The heat flux $\hat{Q}_i$ for the self-consistent value of flow shear $\omega^{\Pi_i=0}_{\perp}$ (ensuring $\hat{\Pi}_i=0$) driven by up-down asymmetric flux surfaces (blue), compared to the same cases without any flow shear $\omega_{\perp}=0$ (red) as a function of $R_0/L_{Ti}$ for (a) $q=2.05, \hat{s}=0.1$, (b) $q=2.05, \hat{s}=0.4$, (c) $q=2.05, \hat{s}=0.8$, and (d) $q=1.05, \hat{s}=0.8$. The Prandtl numbers (black) are also shown for each equilibrium. The flow shear values $\omega^{\Pi_i=0}_{\perp}$ (unit $c_s/R_0$) required to achieve $\hat{\Pi}_i=0$ are indicated by the numbers neighboring each blue data point. \textcolor{black}{The black dashed line in (d) denotes the critical gradient of the $\omega_{\perp}=0$ cases using a linear fitting.}}
    \label{fig2_Reduction}
\end{figure}

\begin{figure}
    \centering
    \includegraphics[width=0.94\textwidth]{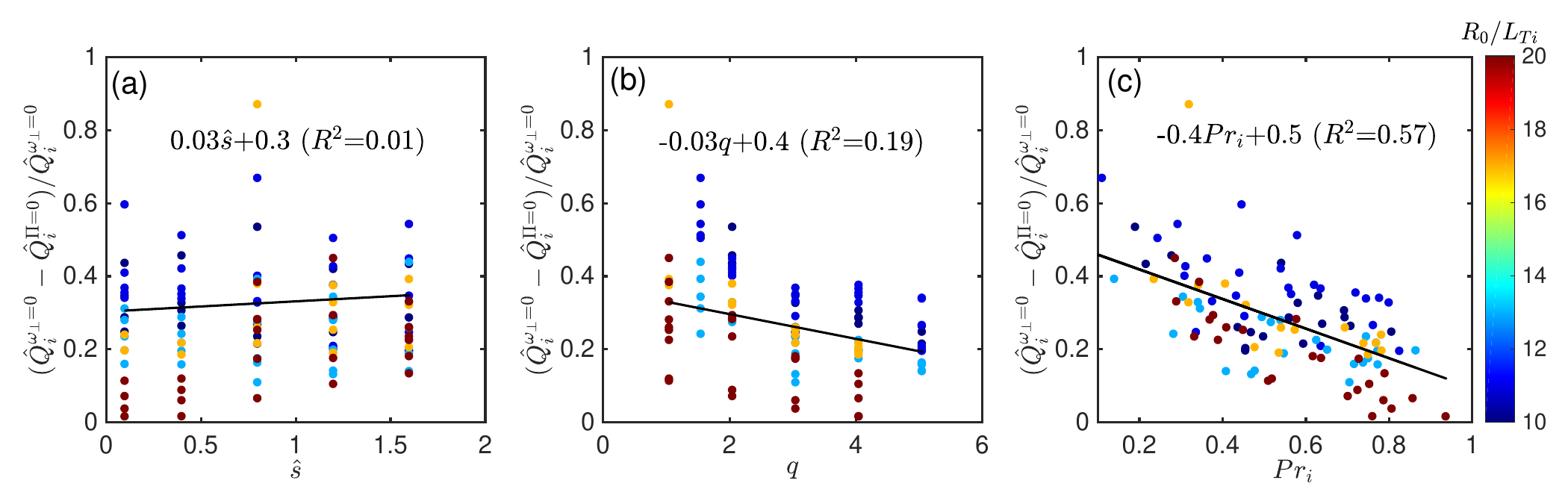}
    \caption{\textcolor{black}{The fractional heat flux reduction defined as $(\hat{Q}^{\omega_{\perp}=0}_i-\hat{Q}^{\Pi=0}_i)/\hat{Q}^{\omega_{\perp}=0}_i$ as a function of (a) magnetic shear, (b) safety factor, and (c) Prandtl number for all the simulation cases that we have run. The colormap on the data points denotes the temperature gradient. The black lines represent the lines of best fit for all the data. Their formulas and $R^2$ values are written on the plot.}}
    \label{fig_Reduction_vs_para}
\end{figure}

Figure \ref{fig2_Reduction} summarizes the effectiveness of up-down asymmetry combined with the LMD regime to create flow shear that stabilizes turbulence. \textcolor{black}{In these NL simulations, both external flow shear and self-generated zonal flow are presented and they both can play a role in saturating turbulence.}
When going from subplot (a), (b), (c) to (d), one gradually moves into the LMD regime. As we can see, the intrinsic flow shear strengthens in the LMD regime, together with the decrease in Prandtl number, enabling a reduction in the heat flux 
by the intrinsic flow shear. Please note that, while momentum diffusivity plays an important part, flow shear stabilization is a combined effect of many different physical processes. For example, in addition to lowering the Prandtl number, higher magnetic shear also strengthens the intrinsic momentum flux (see Fig. \ref{fig2_flowshearscan} (a)) and enables a given amount of flow shear to more effectively reduce the heat flux \cite{Roach_2009}. 
Outside the LMD regime ($\overline{\text{Pr}}_i\gtrsim 0.5$ and low magnetic shear), illustrated by Figs. \ref{fig2_Reduction} (a) and (b), only weak flow shear ($|\omega_{\perp}|\lesssim 0.2c_s/R_0$) is produced and the heat flux is barely affected. In the LMD regime ($\overline{\text{Pr}}_i\lesssim 0.5$ and high magnetic shear), shown in Fig. \ref{fig2_Reduction} (c) and (d), the Prandtl number is low, enabling a strong intrinsic flow shear ($|\omega_{\perp}|\gtrsim 0.3c_s/R_0$) that significantly reduces the heat flux and increases the critical gradient by up to $25\%$ \textcolor{black}{(increasing the critical $R_0/L_{Ti}$ from $13.2$ to $16.9$ in Fig. \ref{fig2_Reduction} (d)).}
Therefore, a combination of LMD with up-down asymmetry can create strong stabilizing intrinsic flow shear in the tokamak core. \textcolor{black}{To show that the Prandtl number is a key parameter for enabling the turbulence stabilization, we plot in Fig. \ref{fig_Reduction_vs_para} the fractional heat flux reduction $(\hat{Q}^{\omega_{\perp}=0}_i-\hat{Q}^{\Pi=0}_i)/\hat{Q}^{\omega_{\perp}=0}_i$ as a function of various parameters for all the cases that we run. Here $\hat{Q}^{\omega_{\perp}=0}_i$ and $\hat{Q}^{\Pi=0}_i$ correspond to the red and blue lines in Fig. \ref{fig2_Reduction}, respectively. The parameters that we consider are $\hat{s}$, $q$ and $\text{Pr}_i$, as they all strongly affect the heat flux reduction. The color scheme of the data points represents the ion temperature gradient, showing that a lower temperature gradient (closer to marginal stability) generally leads to more significant heat flux reduction. As can be seen in Fig. \ref{fig_Reduction_vs_para} (a) and (b), the data points are strongly scattered when plotting against $\hat{s}$ or $q$. When the data is fit with a line, it has $R^2$ values of $R^2=0.01$ and $R^2=0.19$. This indicates that the heat flux reduction depends on more than one of these basic parameters. However, according to Fig. \ref{fig_Reduction_vs_para} (c), the fractional heat flux reduction shows a significantly better correlation
(linear fit with $R^2=0.57$) with respect to the Prandtl number $\text{Pr}_i$. At a Prandtl number of $\text{Pr}_i=0.5$ (the boundary of the LMD regime), the average heat flux reduction is around 30\% based on the linear fit in Fig. \ref{fig_Reduction_vs_para} (c). Therefore, the Prandtl number appears as the most effective parameter in determining the fractional heat flux reduction. A low Prandtl number increases the fractional heat flux reduction and helps stabilize turbulence.}


\begin{figure}
    \includegraphics[width=0.94\textwidth]{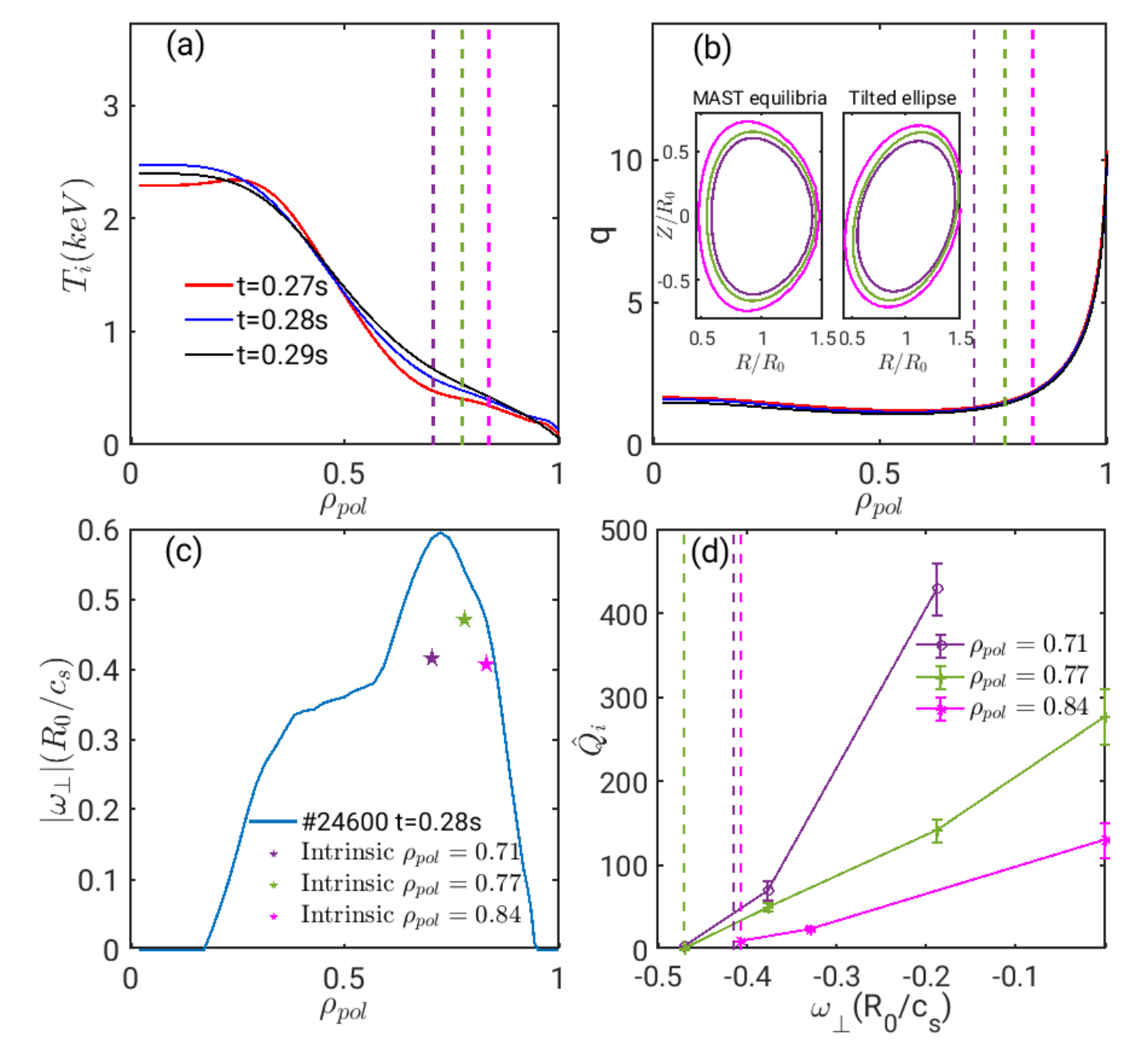}
    \centering
    \caption{Subplots (a) and (b) show the radial profiles of ion temperature and safety factor measured for MAST shot 24600 at $t=0.28s$. The profiles are obtained from TRANSP interpretative analysis, using the data processing code Pyrokinetics \cite{Patel_pyrokinetics_2022}. The three vertical dashed lines indicate the three considered radial locations $\rho_{pol}=\left\{0.71, 0.77, 0.84\right\}$. The inset in (b) shows the MAST experimental flux surface shape at these three radial locations as well as the artificially tilted elliptical geometry. (c) shows the experimentally measured toroidal flow shear $|\omega_{\perp}|$ produced by  NBI (blue line) and the GENE predictions for intrinsic flow shear (stars) that could be created intrinsically if MAST could create the tilted elliptical geometries. (d) shows the heat flux with flow shear using the tilted elliptical geometry, where the vertical dashed lines indicate the predicted steady-state intrinsic flow shear.}
    \label{fig3_profile}
\end{figure}

To check the experimental feasibility of this strategy, we take an experimental equilibrium from the Mega Ampere Spherical Tokamak (MAST). \textcolor{black}{We consider shot 24600 at time $t=0.28s$. This is an L-mode plasma with 3MW of on-axis co-NBI injection starting from $t=0.2s$, creating strong flow shear and driving momentum flux \cite{Field_2011NF}. The ion temperature profile is shown in Fig. \ref{fig3_profile} (a) and the flux surfaces and $q$ profile are shown in Fig. \ref{fig3_profile} (b).} The experimentally measured kinetic profiles are averaged over $0.27s\leq t\leq 0.29s$ to reduce the numerical uncertainties. We use $\rho_{pol}=\psi^{1/2}_n$ as the radial coordinate, where $\psi_n$ is the normalized poloidal flux. As we can see, the $q$ profile is low over a wide radial range, which, together with the tight aspect ratio of MAST, facilitates LMD. No significant MHD instabilities are observed at around $t=0.28s$ \cite{Field_2011NF} and the system is in a quasi-steady state. We perform gyrokinetic simulations at three radial locations $\rho_{pol}=\left\{0.71, 0.77, 0.84\right\}$, which are dominated by ITG driven turbulence. We tested other radial locations, but they are dominated by either Micro-Tearing Modes (MTM) or Electron Temperature Gradient (ETG) modes, which are computationally much more challenging for NL gyrokinetic simulations. 
Kinetic electrons are included in these simulations. Our simulations indicate that these three radial locations are close to or in the LMD regime, with $\text{Pr}_{i}=\left\{0.58, 0.40, 0.41\right\}$. 

To verify that one can expect the LMD regime to reduce turbulence, we artificially modified the MAST flux surfaces to be tilted ellipses as shown in Fig. \ref{fig3_profile} (b) without considering the external torque from NBI, and performed NL flow shear scans to find $\omega^{\Pi_i=0}_{\perp}$. These flow shear values are shown in Fig. \ref{fig3_profile} (c), alongside the experimental flow shear profile. This illustrates that experimentally significant flow shear levels driven by external NBI sources could actually be driven intrinsically if MAST could achieve tilted elliptical flux surfaces. Figure \ref{fig3_profile} (d) shows that this intrinsic flow shear almost entirely quenches the ITG turbulence and thus significantly reduces the heat flux. \textcolor{black}{As shown in our companion paper \cite{sun2024physicslowmomentumdiffusivity}, this heat flux reduction by the intrinsic flow shear is comparable to the experimental heat flux with strong NBI.}

\begin{figure}
    \includegraphics[width=0.94\textwidth]{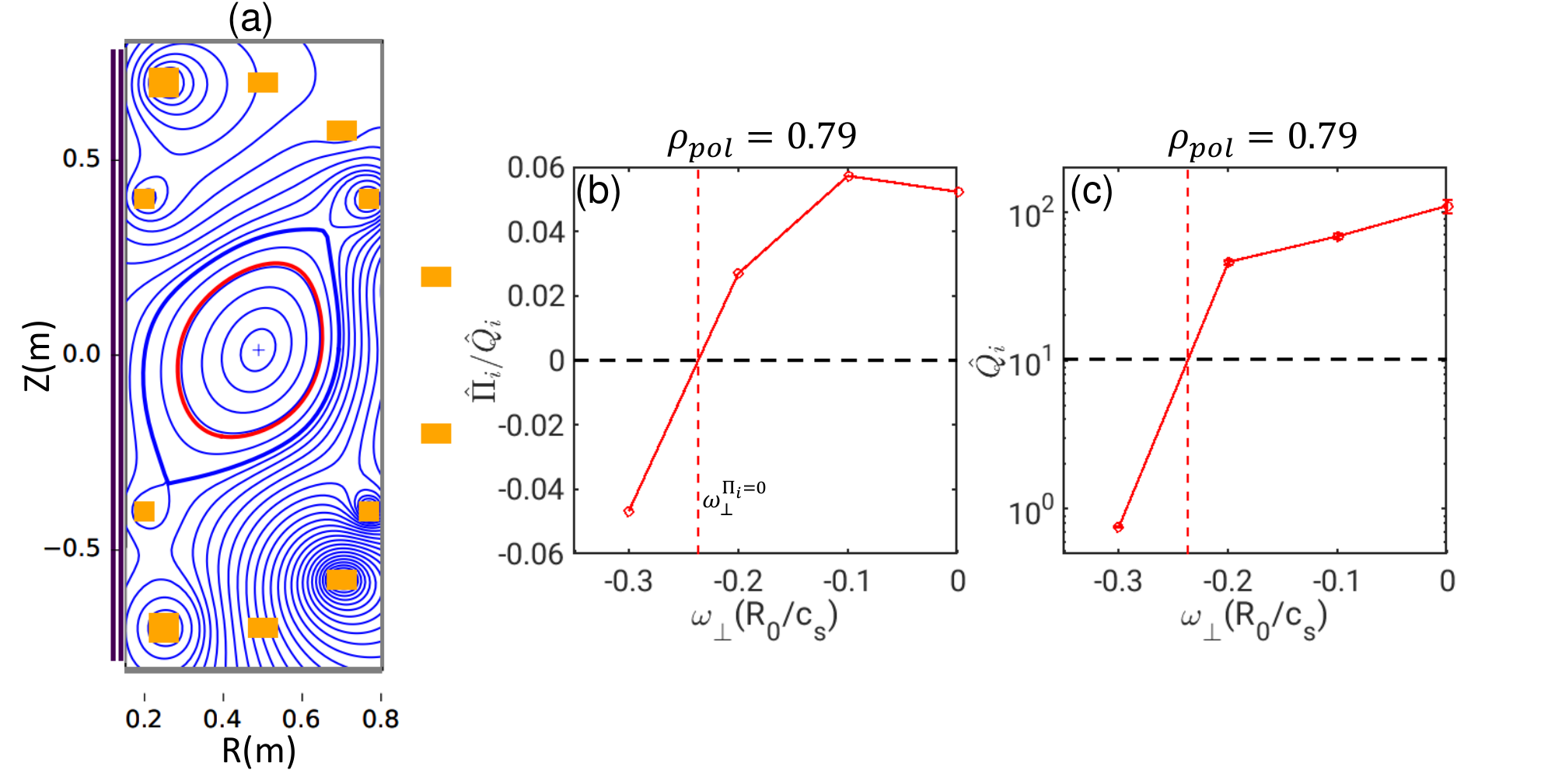}
    \centering
    \caption{Subplot (a) shows an up-down asymmetric equilibrium that can be achieved by the SMART tokamak with its advanced coil set. Subplots (b) and (c) show the ratio $\hat{\Pi}_i/\hat{Q}_i$ and $\hat{Q}_i$ as functions of flow shear from NL GENE simulations at $\rho_{pol}=0.79$.}
    \label{fig5_SMARTstudy}
\end{figure}

While it is difficult to create strongly up-down asymmetric flux surfaces in MAST, a new spherical tokamak under construction, SMART \cite{DOYLE_SMART_2021,MANCINI_SMART_2021,AGREDANOTORRES_SMART_2021,SEGADOFERNANDE_SMART_2023,Podesta_SMART_2024}, has stronger shaping capabilities. Figure \ref{fig5_SMARTstudy} (a) shows a preliminary up-down asymmetric geometry that can potentially be achieved in SMART with the addition of its advanced coils. We consider one radial location at $\rho_{pol}=0.79$, dominated by ITG turbulence, and perform a similar NL flow shear scan as we did for MAST. The important parameters are $R_0/L_{Ti}=6.32$, $q=1.34$, $\hat{s}=1.25$, $\epsilon=0.39$. A strong intrinsic flow shear of $|\omega^{\Pi_i=0}_{\perp}|=0.23c_s/R_0$, together with one order of magnitude heat flux reduction (shown in Fig. \ref{fig5_SMARTstudy} (b) and (c), respectively) are observed, further demonstrating the experimental significance of our method.

\textbf{\textit{Conclusions--}}In this letter, we propose and analyze a new method to drive strong intrinsic rotation shear in tokamaks: combining the LMD regime with up-down asymmetry. Using a large number of NL and QL gyrokinetic simulations, we self-consistently calculate the flow shear and show that it can significantly stabilize ITG turbulence. \textcolor{black}{This is facilitated by high $\hat{s}$, low $q$, tight aspect ratio, and temperature gradient close to marginal stability.} The experimental feasibility of this method is addressed by considering a MAST experimental equilibrium and a planned SMART equilibrium. 
Moreover, our results actually present a lower bound for the impact of the intrinsic flow shear as we neglected the pinch term in our analysis \cite{sun2024physicslowmomentumdiffusivity} and any external momentum drive. Based on our preliminary study, SMART should be able to test the main predictions in this letter. 
This novel approach for generating strong flow shear does not rely on external momentum injection and therefore could be an attractive way of creating strong rotation in future large spherical tokamaks like STEP \cite{ANAND_STEPstatus}.

\textbf{\textit{Acknowledgement--}}The simulations in this work are performed on CSCS Daint, Cineca Marconi and CSCS Lumi clusters, with a total computational cost of about $3\times 10^7$ CPU hours. The authors thank Prof. Ben McMillan, Dr. Antoine Hoffmann, Dr. Francis Casson, Arnas Volcokas and Alessandro Balestri for the fruitful discussions. This work was supported by a grant from the Swiss National Supercomputing Centre (CSCS) under project ID s1097. This work has been carried out within the framework of the EUROfusion Consortium, partially funded by the European Union via the Euratom Research and Training Programme (Grant Agreement No. 101052200 - EUROfusion). The Swiss contribution to this work has been funded by the Swiss State Secretariat for Education, Research and Innovation (SERI). Views and opinions expressed are however those of the author(s) only and do not necessarily reflect those of the European Union, the European Commission or SERI. Neither the European Union nor the European Commission nor SERI can be held responsible for them. This work was supported in part by the Swiss National Science Foundation and by the EPSRC Energy Programme (Grant Number EP/W006839/1).


\section*{References}
\bibliography{apssamp}

\end{document}